\documentclass[twocolumn,aps,pre,floatfix,footinbib]{revtex4}

\usepackage{graphicx,color,amsmath}

\newcommand{\figpntJ}{%
\begin{figure}[b]
   \includegraphics[width=1.in,clip]{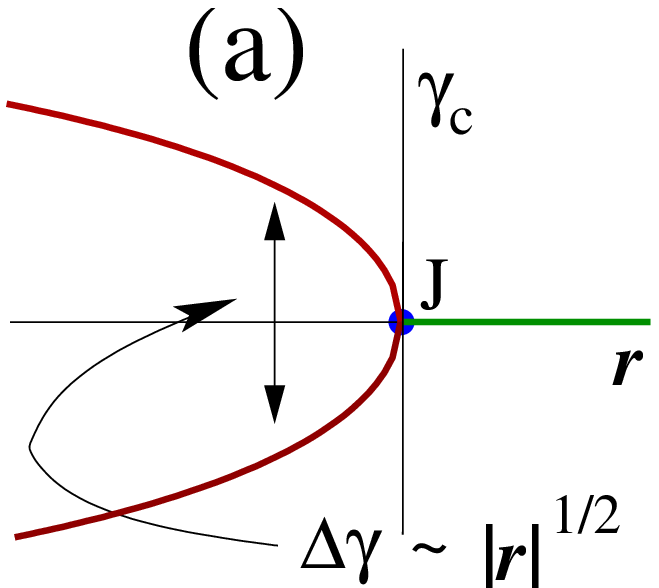} 
    \includegraphics[width=2.in,clip]{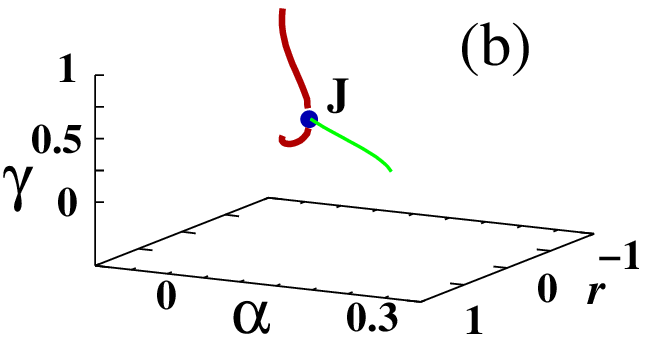} 
   \caption{(a) The critical lines in the
   $\gamma$-$r$ plane, where $r=\varepsilon-\varepsilon_J$. The width
   vanishes with a characteristic exponent as $r\to 0-$.
 (b) Three
   dimensional view of the tricritical point J in the
   $\alpha$-$\gamma$-$r$ space. Only the critical lines are
    shown. These lines are nonplanar and (a) shows the projection.}
\label{fig:pntJ}
\end{figure}
}

\newcommand{\figjrho}{%
\begin{figure}[t]
   \includegraphics[height=1in,clip]{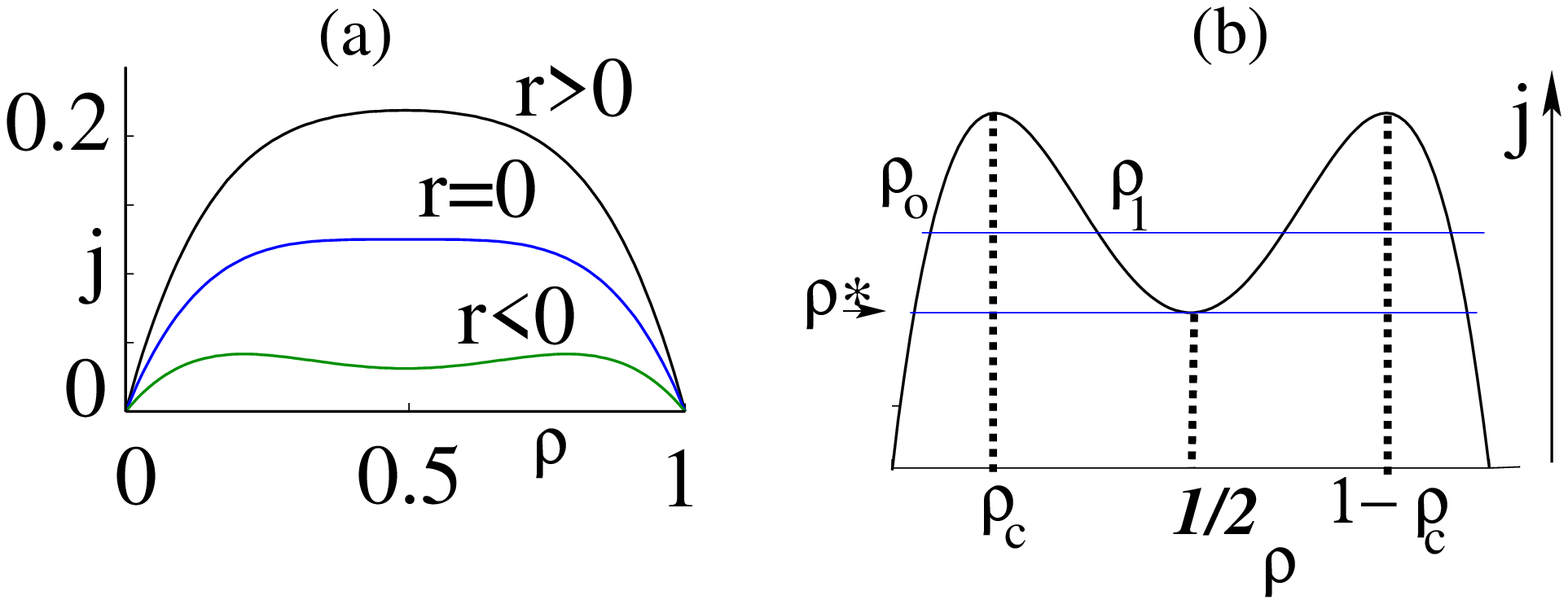} 
   \caption{(a)
     Current versus $\rho$ for three values of $r$.  (b) shows the
     notations used for $r<0$.  The peak densities are $\rho_c$ and
     $1-\rho_c$.  The inner layer connects pairs of densities (with
     same current) from $\rho_o, \rho_1, 1-\rho_1,1-\rho_o$.  For
     $\rho_1=1/2$, the low and high densities are $\rho^*$ and
     $1-\rho^*$.  For currents les than $j(1/2)$ (and also for $r\geq
     0$), $\rho_o$ and $1-\rho_o$ are the two relevant densities.  }
\label{fig:jrho}
\end{figure}
}

\newcommand{\figdel}{%
\begin{figure}[t]
   \includegraphics[width=3.in,clip]{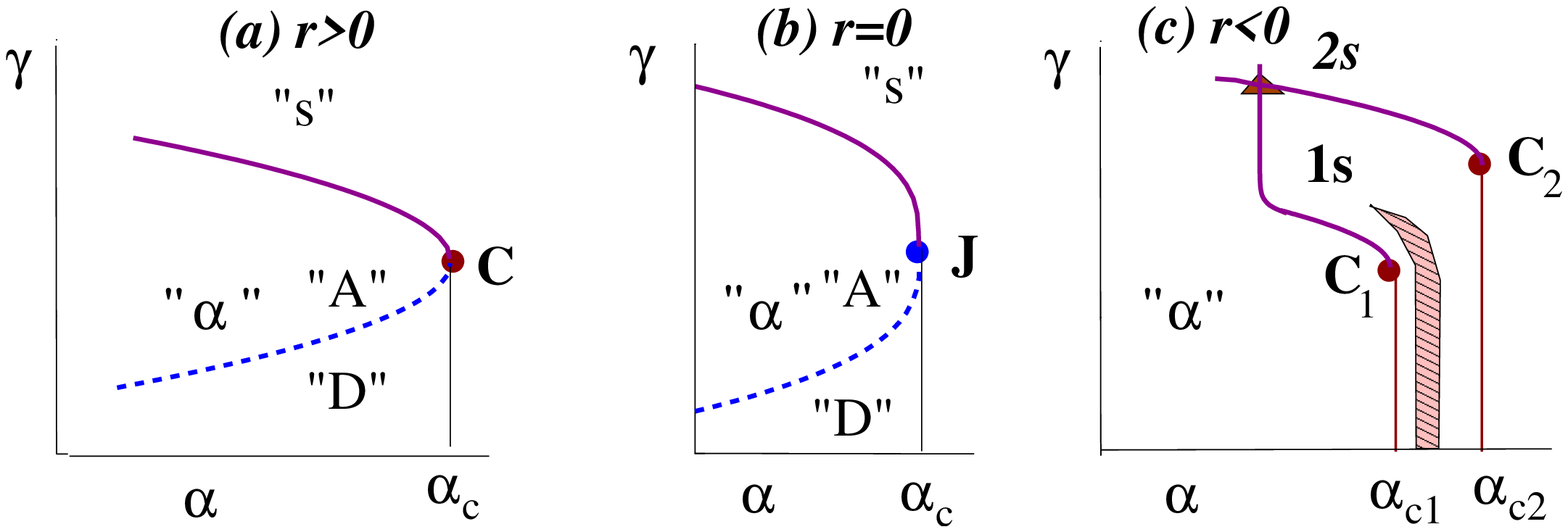} 
   \caption{Schematic diagram of the phase boundary for shock at
     $x=1$.  Thin, thick and dotted lines represent continuous
     transitions, first order transitions and dual lines respectively.
     ``A''[``D'']: accumulated [depleted] bounary layer. C, C$_1$,
     C$_2$ are the critical points and J the tricritical point.
     ``$\alpha$'' and ``s'' represent $\alpha$-phase and shock phase
     with 1,2 denoting 1-shock, 2-shock phases.  The phase boundary
     separating the $\alpha$ and the 1s phases is given by Eq.
     (\ref{eq:22}).  Downward shocks appear in the hatched region
     (with $\gamma\le 1/2$) in (c).  The dual lines demarcating
     ``A"/``D" are not shown in (c).}
\label{fig:del}
\end{figure}
}

\newcommand{\figrhox}{%
\begin{figure}[b]
   \includegraphics[width=1.5in,clip]{frst_crt_2.eps}
   \includegraphics[width=1.5in,clip]{2nd_crt_2.eps} 
   \caption{ Plots of $\rho(x)$ vs. $x$ ($r=-0.2,u=2.2,\Omega=0.1$)
     from numerical solution.  The values of $\gamma$ are indicated on
     the plots while $\alpha$ is kept fixed, close to $\alpha_{c1}$ in
     (a) and $\alpha_{c2}$ in (b). In (a), as $\gamma$ increases, one
     sees a sequence: (i) the shock developing from zero height
     ($\gamma=0.35$) to a maximum ($\gamma=0.39$), (ii) an additional
     downward shock ($\gamma=0.41$) which goes towards the boundary
     ($\gamma=0.56$), (iii) the emergence of another upward shock of
     finite height ($\gamma=0.70$) , and (iv) finally a bigger shock
     symmetric around $\rho=1/2$ ($\gamma=0.71$).  (b) shows the
     sequence for the second critical point. There is no downward
     shock here, while the first shock is of maximum possible height.
     To be noted is the depletion layer for $\gamma=0.62$.  }
\label{fig:rhox}
\end{figure}
}

\newcommand{\upd}{{\rm d}}
\begin{document}

\title{Nonequilibrium tricriticality in one dimension} 

\author{Jaya Maji and Somendra M.  Bhattacharjee}
 
\affiliation{Institute of Physics, Bhubaneswar-751005, India }

\begin{abstract}
{{We show the existence of a nonequilibrium
    tricritical point induced by a repulsive interaction in one
    dimensional asymmetric exclusion processes.}  The tricritical
  point is associated with the particle-hole symmetry breaking
  introduced by the repulsion.  The phase diagram and the crossover in
  the neighbourhood of the tricritical point for {the shock
    formation} at one of the boundaries are determined.}
\end{abstract}

\maketitle

The lack of any free energy like entity to describe nonequilibrium
steady states makes general studies of nonequilibrium phase
transitions rather difficult.  For this reason, eventhough different
types of phase diagrams are known with first-order or continuous phase
transitions, and critical points from various case
studies\cite{derrida1,frey,evans,smsmb,smvm}, generalizations of these
results are not straightforward.  Since steady states are the closest
analogs of equilibrium {states (both being stationary in
  time)}, it is important to know how far the
richness of the equilibrium critical phenomena with their connections
to symmetries\cite{huang} can be found in nonequilibrium systems.

This paper shows the bifurcation of a line of critical points through a
tricritical point\cite{huang,ahar} in the phase diagram of a one
dimensional interacting driven system as one interaction parameter is
tuned.  A tricritical point is also a point where a first order
transition changes over to a continuous one, the pair of critical
lines being the edges of two first order surfaces.  We do observe such
equilibrium-like gross features in the nonequilibrium system so as to
classify it as a tricritical point, but, unlike equilibrium systems,
all happening {\it in one dimension}\footnote{ Though nonequilibrium
  tricritical points in dimensions $>1$ are known\cite{comm}, none of
  these would show tricriticality in one-dimension.}.

The occurrence of a tricritical point in any system is significant
because it implies a confluence of two different phenomena.  As a
consequence, the tricritical point controls the scaling description in
its neighbourhood.  Known examples from equilibrium systems include
the $\theta$-point of polymer solutions\cite{degen}, bunching and
phase separation of steps on crystal surfaces\cite{smbsi}, different
transitions in magnets\cite{ahar}, current hunt for a tricritical
point in quark-gluon plasma in the context of early
universe\cite{qgp}, and many others\cite{comm}.  In equilibrium
systems, a tricritical point is one in the hierarchy of critical
points and corresponds to the case of three relevant variables as
opposed to two for an ordinary critical point, as e.g., temperature
and magnetic field for a Curie point of a magnet.  In the Landau
theory of phase transitions, the sequence of critical, tricritical
points ... occur as new symmetric minima in the Landau function
develop.  In analogy with that, we find in the nonequilibrium problem
in hand, the tricriticality is associated with a particle-hole
symmetry breaking or successive occurrences of peaks in the current in
the system, that allows maximum current through the system at two
different densities.

\figpntJ

Our results are based on a class of well-studied one-dimensional
models for interacting particles moving on a track, which are variants
of the asymmetric exclusion process (ASEP).  There is current interest
in these models because of the recent observation of localized shock
phases and associated criticality. A shock in these models is a
discontinuity in the steady state density along the track.  Let us
consider a one-dimensional lattice with particles injected at site
$i=0$ at a rate $\alpha$ and withdrawn at $i=N$ at a rate $1-\gamma$.
The particles hop to the right with mutual exclusion so that the
occupation number at a site is $0$ or $1$.  In addition, there is a
next nearest neighbour repulsion\cite{hager} so that a configuration
\begin{eqnarray*}
  1100 &\to& 1010 \qquad {\rm with\  a \ rate\ } 1+\varepsilon,\\ 
  0101 &\to& 0011 \qquad {\rm at \ a \ rate\ } 1-\varepsilon,
\end{eqnarray*}
with ($0<\varepsilon <1$), where occupied and unoccupied states of a
site are represented by $1$ and $0$ respectively.  The hopping process
is particle conserving. { With a nonvanishing current, the
  system can evolve to a nonequilibrium steady state.}
Nonconservation is introduced by allowing evaporation or desorption
($1\to 0$) at a rate $\omega_d$ and deposition or adsorption ($0\to
1$) at a rate $\omega_a$ of particles at any site on the track.  This
dynamics, called the Langmuir dynamics, can maintain a density
\begin{equation}
  \label{eq:5}
  \rho_L=\frac{\omega_a}{\omega_a+\omega_d} \neq \frac{1}{2} \quad
  {\rm if}\  \omega_a\neq\omega_d.
\end{equation}
{This is an equilibrium-like process with no current in the
  system.  If the net flux due to adsorption/desorption is comparable
  to the hopping current, there will be a competition between the
  attempt to equilibrate and the drive to the nonequilibrium steady
  state.  As a result, one gets new features like a shock phase,
  critical points one would not get otherwise.}

For large $N$, we may use a continuum notation $x=i/N$ and the average
occupation of a site becomes the density $\rho(x)$.  For given
microscopic parameters for the hopping rules and nonconservation, the
steady-state density profile $\rho(x)$ depends on the external
parameters $\alpha$ and $\gamma$.  The possible nonequilibrium phases
are then represented in a phase diagram in the
$\alpha$-$\gamma$-$\varepsilon$ space.

The extra interaction $\varepsilon$ is important in particle-hole
symmetry breaking.  To see this let us consider the conserved case
($\omega_a=\omega_d=0$).  Instead of particle injection at $x=0$, we
may consider holes being injected at the right ($x=1$) at a rate
$1-\gamma$, hopping left with identical rates and withdrawn at $x=0$
at a rate $\alpha$.  This particle-hole symmetry allows a symmetric
form for current $j(\rho)=j(1-\rho)$.  For $\varepsilon=0$,
$j(\rho)=(1-\rho)\rho$ satisfies the symmetry with a maximum at
$\rho=1/2$.  A similar single peaked current with $j(\rho)=j(1-\rho)$
is expected for low interaction strengths (small $\varepsilon$).  If a
system evolves to the maximal current state then the steady state or
the phase also has the particle-hole symmetry ($0\leftrightarrow 1$).
In addition to this symmetric phase, there are two distinct phases
related by the $0\leftrightarrow 1$ symmetry but separated in the
phase diagram by a first order line.  In the other extreme case
$\varepsilon=1$ which forbids the hops $0101 \to 0011$, there is a
zero current state for $\rho=1/2$ as for $\rho=0$ (empty track) and
$\rho=1$ (packed track).  Therefore, for strong enough interaction,
$\rho=1/2$ is a local minimum of the current while the maximal current
carrying states are symmetrically located on the two sides of
$\rho=1/2$.  The exactly known stationary current for the conserved
system does show the change from single to double peak for
$\varepsilon$ close to $\varepsilon=\varepsilon_{\rm J}\approx
0.8$\cite{krug91,hager}.  In any one of the maximal current state, the
particle-hole symmetry is not respected\footnote{ This is to be
  distinguished from explicit breaking by other interactions that make
  the current asymmetric with $j(\rho)\neq j(1-\rho)$.}.  The overall
symmetry is recovered by recognizing that the $0\leftrightarrow 1$
transformation gives the other maximal current state.  The single
maximal current phase in the single-peak case now breaks up into
several phases, two of which are the two maximal current phases.  The
occurrence of the two maximal current phases in different regions of
the phase diagram is a reflection of the symmetry breaking introduced
by the interaction.

For the ASEP case (i.e., with $\varepsilon=0$ in the above
model)\cite{frey,evans,smsmb}, for $\omega_a\neq\omega_d$, there are
three types of phases, namely, (i) an injection rate controlled, to be
called the $\alpha$-phase, (ii) a withdrawal rate controlled phase, to
be called the $\gamma$-phase, and, most importantly, (iii) a phase
containing a localized shock, to be called a shock phase.  The maximal
current phase of the conserved case becomes unstable under
nonconservation, except for the case of accidental symmetry at
$\omega_a=\omega_d$.  The major signature of nonconservation is in the
appearance of the shock phase with a localized shock in the density
profile in between the $\alpha$- and the $\gamma$-phases.  This phase
replaces the first-order boundary of the conserved case.  The maximal
current density is still important because the shock is centered
around the maximal current density.  In other words, a shock connects
configurations related by the particle hole symmetry, i.e., density
profiles with $\rho=1/2\pm\delta$.  The transition to the shock phase
is generally first order because the shock height is nonzero at the
transition, but critical points do occur where the shock height
vanishes.  The vanishing height implies a special state that regains
the particle-hole symmetry.  Such a critical point therefore has to be
at the peak of the current.  There are characteristic universal
exponents, for both bulk\cite{frey} and boundary\cite{smvm},
associated with the critical point.

In the interacting case ($\varepsilon> 0$) with conservation
($\omega_a=\omega_d=0$), a very complex phase diagram with seven
different phases including two maximal current phases (mentioned
earlier) is known but {\it without any shock phase}, let alone the
critical point.  The dynamics of the interacting model has also been
studied in certain regions of the parameter space in Refs.
\cite{hager,popkov}.  With nonconservation, localized shocks appear,
and double shocks and downward shocks have also been
seen\cite{popkov,iran,smdn} but the phase diagram is not known.  In
the broken symmetry case, a shock centered around one peak will not
show the $0\leftrightarrow 1$ symmetry and in fact the state obtained
by such transformation would occur in a different region of the
parameter space.  Given the fact that the maximal current densities
play an important role in shock formation, the symmetry-breaking is
expected to show new critical points compared to the symmetric case.
Consequently, the critical point for the shock phase for
$\varepsilon<\varepsilon_{\rm J}$ appears as a line in the extended
$\alpha$-$\gamma$-$\varepsilon$ phase diagram and this line undergoes
a bifurcation at $\varepsilon=\varepsilon_J$ where the symmetry
breaking takes place.  The bifurcation point J is the tricritical
point as shown in Fig.  \ref{fig:pntJ}, and it has its own distinct
scaling.

The phase transitions can be analyzed by using a boundary layer
analysis for the density profile in the continuum, long time, long
length scale limit, the so-called hydrodynamic regime
\cite{smsmb,smb_jpa07}.  The shock phase can be seen as forming via a
thickening and eventual deconfinement of a boundary layer (`shockening
transition').  Moreover, this bulk cum boundary shockening transition
is in turn associated with a dual boundary transition where the
boundary layer changes from a depletion region to an accumulated
region \cite{smvm,smb_jpa07}.  If the shockening and the dual
transition lines intersect then there is a critical point
(``self-dual'').

So far as the single to double peak change in the current is
concerned, a Taylor series expansion in $\rho$ and $\varepsilon$ of
the exactly known current\cite{hager} upto fourth order is sufficient.
For simplicity we work with this expansion in a general form
\begin{equation}
  \label{eq:3}
  j(\rho)= \frac{2r+u}{16} - \frac{r}{2} (\rho-\frac{1}{2})^2 
  - u\  (\rho-\frac{1}{2})^4,
\end{equation}
with the constant piece chosen to ensure that the current vanishes for
$\rho=0,1$ (the empty track and the fully-occupied track).  Eq.
\eqref{eq:3} for $r=2, u=0$ recovers the known current density for
ASEP processes ($j=(1-\rho)\rho$) while the double peak appears for
$r<0$.  Throughout $u$ is taken as a positive constant, and $r$ is
tunable but kept small.  The current is shown in Fig.  \ref{fig:jrho}a
for typical values of $r$.  In the double peak case, there are a few
special densities we need, namely, (i) the densities for maximal
current, $\rho_c, 1-\rho_c$, and (ii) the densities $\rho^*,
1-\rho^*$, with the same current as the minimal one $(j(1/2))$.  Any
constant current line with $j(\rho_c)>j>j(1/2)$ intersects at four
densities $\rho_o, \rho_1, 1-\rho_1, 1-\rho_o$.  There are only two
values $\rho_o, 1-\rho_o$ if $j< j(1/2)$. These are shown in Fig.
\ref{fig:jrho}b.

\figjrho
 
The hydrodynamic approach based on the continuity equation with
$j(\rho)$ of the conserved case is known to agree remarkably with
numerical simulations\cite{hager,popkov}.  {In this approach
  the density satisfies a continuity equation supplemented by an
  additional contribution from the nonconserving process.  The density
  then satisfies
\begin{equation}
  \label{eq:24}
  \frac{\partial \rho}{\partial t} + \frac{\partial J}{\partial x} + S_0=0,
\end{equation}
where $J$ is the current consisting of the bulk current $j(\rho)$ and
a diffusive current, and 
\begin{equation}
  \label{eq:25}
   S_0(\rho)=-\Omega (\rho_{\rm L}-\rho),\quad \Omega=(\omega_a+\omega_d) N,
\end{equation}
accounts for the Langmuir dynamics.  $\Omega$ is related
to the net flux of particles and $\rho_{\rm L}$ is the density defined
in Eq.  \ref{eq:5}.  Nonconservation would matter only when the net
flux of particles adsorbed or desorbed is comparable to the current in
the system.  For this reason $\Omega$ is kept constant, finite
(scaling limit) for $N\to\infty$.  By adding the  diffusive (Fick's
law) current,  $J$ in Eq. \ref{eq:24} can be written as 
\begin{equation}
  \label{eq:26}
  J= -\epsilon \frac{\partial \rho}{\partial x} + j(\rho),
\end{equation}
with the diffusion constant $\epsilon\sim O(1/N)$ is a small parameter
(not to be confused with $\varepsilon$), and is a remnant of the
underlying lattice.  This form, Eq. \ref{eq:26}, also follows from a
continuum limit of the discrete model. Despite its smallness
$\epsilon$ plays an important role in the phase transitions,
especially in shock formation.  The steady state density profile is
now given by
\begin{subequations}
\begin{eqnarray}
  \label{eq:4}
  &&    -\epsilon  \frac{\upd^2 \rho}{\upd x^2} 
  +S_1(\rho)\frac{\upd \rho}{\upd x} + S_0(\rho)=0,\\
  {\rm where}&&
  S_1(\rho)=\frac{\upd j(\rho)}{\upd\rho}, \label{eq:4c}
\end{eqnarray}
\end{subequations}
and the boundary conditions $\rho(0)=\alpha$, and $\rho(1)=\gamma$.
We already noted that $\rho=1/2$ is a special density due to the
particle-hole symmetry.  To avoid extra complications arising from the
accidental matching of densities, we take $\omega_a \neq \omega_d$ so
that $\rho_{\rm L}\neq 1/2$.  Such cases will be considered
elsewhere.}

Eq. \ref{eq:4} entails two length scales, (i) $x$ for the bulk and
(ii) $\tilde{x} = (x-x_0)/\epsilon$ which is significant in a thin
region as $\epsilon \to 0$ around an appropriately chosen $x_0$. Eq.
\ref{eq:4} does not depend on the choice of $x_0$.  The separation of
the two scales is used to develop a uniform approximation of the
solution order by order in $\epsilon$. Here we restrict ourselves to
the lowest order approximant.  The bulk solution in terms of $x$ comes
from the first order equation obtained with $\epsilon=0$ in Eq.
\ref{eq:4}.  This solution, to be called the outer solution,
\begin{equation}
  \label{eq:5n}
 \rho(x)=\rho_{\rm out}(x) 
\end{equation}
is given implicitly by
\begin{subequations}
  \begin{equation}
    \label{eq:17}
    g_r(\rho) = 2 \Omega \ x + C,
   \end{equation}
where $C$ is a constant to be fixed by only one of the two boundary
conditions, and
\begin{eqnarray}
 \label{eq:6}
\hspace*{-.5cm}g_r(\rho)  &\equiv& (u Y^2_{\rm L}+ r)\:  Y 
          + \frac{u}{2} Y_{\rm L} Y^2 +
\frac{u}{3} Y^3 \nonumber\\
              &&  \qquad + (u Y^2_{\rm L}+ r)Y_{\rm L} \ln \mid Y_{\rm L}-
                   Y\mid
\end{eqnarray}
\end{subequations}
with $Y=2\rho -1$,and $Y_L=2\rho_L-1$. We shall suppress the
$r$-dependence in $g_r$, unless required.  For the left boundary
condition, the outer solution is
\begin{equation}
\label{eq:7}
g(\rho)=2 \Omega \ x +g(\alpha),
\end{equation}
with a density 
\begin{equation}
  \label{eq:20}
  \rho_o\equiv \rho_{\rm out}(1) \ {\rm at} \ x=1,
\end{equation}
determined by
\begin{equation}
  \label{eq:15} 
     g(\rho_o)=2\Omega+ g(\alpha).
\end{equation}
Here, $\rho_o$ depends on $\alpha$, but in general, $\rho_o\neq
\gamma$.

To satisfy the other boundary condition $\rho=\gamma$ at $x=1$, we use
the second scale ${\tilde{x}} =(x-1)/\epsilon$ around the boundary
point $x=1$.  With this variable, the density (to be called the inner
solution) satisfies
\begin{equation}
  \label{eq:8}
 - \frac{\upd^2 \rho_{\rm in}}{\upd \tilde{x}^2}+S_1(\rho_{\rm in})\frac{\upd
   \rho_{\rm in}}{\upd \tilde{x}}=0,
\end{equation}
obtained from Eq. (\ref{eq:4}) by first changing the variable to
${\tilde{x}}$ and then taking $\epsilon\to 0$.  The absence of the
$S_0$ term can be understood from the observation that for a constant
$\Omega$ this layer is too thin for the nonconservation to matter, at
least to leading order in $\epsilon$.  For a smooth density profile
(the solution of the second order equation {with any
  $\epsilon>0$ no matter how small}), we need to match the outer and
the inner solutions by requiring that
\begin{equation}
  \label{eq:16}
  \lim_{x\to 1}\rho_{\rm out}(x)=\lim_{\tilde{x}\to
    -\infty}\rho_{\rm in}(\tilde{x}). 
\end{equation}
Here ${\tilde{x}\to -\infty}$ gives the outer limit of the inner
solution.  Incorporating this matching condition, Eq. \ref{eq:8} can
be written as
\begin{equation}
  \label{eq:1}
  \frac{\upd \rho_{\rm in}}{\upd \tilde{x}} =  j(\rho_{\rm in}) -
j(\rho_o),
\end{equation}
in the thin boundary layer, with $\rho_{\rm in}(\tilde{x}=0)=\gamma$.
The complete matched solution is obtained by joining the inner and
outer approximations and subtracting their common value.  Therefore,
the density profile is given by
\begin{equation}
  \label{eq:7a}
\rho(x)=\rho_{\rm out}(x)-\rho_o+\rho_{\rm in}(\tilde{x})+ O(\epsilon).
\end{equation}
Eq. \ref{eq:7a} identifies the scale dependent separation of the bulk
and the boundary contributions and it provides a uniform approximation
of the density in the whole domain including the boundaries.

\figdel

Let us first consider the $r>0$ case.  The current has only one peak
at $\rho=1/2$.  The transport across the track is analogous to the
well-understood ASEP case.  For a low injection rate $\alpha$, the
bulk density profile does not depend significantly on the withdrawal
rate so long $\gamma$, the accumulated density at $x=1$, is small.
One gets the $\alpha$-phase.  For low $\gamma$, there is a depletion
layer at $x=1$, which changes for $\gamma>\gamma_d\equiv
\rho_o(\alpha)$ to an accumulated layer.  This is the dual boundary
transition mentioned earlier.  By symmetry (Fig. \ref{fig:jrho}) and
from Eq. \eqref{eq:1}, $\rho_{\rm in}(\tilde{x})$ would saturate at
$\rho=1-\rho_o(\alpha)$ as ${\tilde{x}}\to \infty$.  Consequently, for
$\gamma\ge \gamma_s\equiv 1-\rho_o(\alpha)$, a boundary layer given by
Eq. \eqref{eq:1} is not sufficient to satisfy the boundary condition
at $x=1$.  In this situation, the density profile is given by the two
outer solutions,
    $$g(\rho_{\rm rgt})=2\Omega\  (x-1) +g(\gamma)$$ 
on the right side and 
    $$g(\rho_{\rm lft})=2\Omega\ x + g(\alpha)$$
on the left.  These two solutions are joined smoothly in a thin
region by the inner solution of Eq. \eqref{eq:8} with
${\tilde{x}}=(x-x_s)/\epsilon$.  On the bulk scale this looks like
a discontinuity and is therefore a shock at $x=x_s$.  The matching
conditions and the location of the shock ($x_s$) are given by
\begin{subequations}
\begin{eqnarray}
  \label{eq:2}
  \rho_{\rm lft}(x\to x_s-)&=&\lim_{\tilde{x}\to
    -\infty}\rho_{\rm in}(\tilde{x}),\\
 {\rm and}\quad 
  \rho_{\rm rgt}(x\to x_s+)&=&\lim_{\tilde{x}\to
    \infty}\rho_{\rm in}(\tilde{x}),\label{eq:21}
\end{eqnarray}
\end{subequations}
where Eq. \ref{eq:2} is the matching of the left outer solution to the
left end of the inner solution while Eq. \ref{eq:21} is the matching
for the right side.  This deconfinement of the boundary layer by
shifting into the bulk as a shock is the shockening transition.  For a
given $\alpha$, the shock and the dual transitions are given by
\begin{eqnarray}
  \label{eq:18}
  \gamma&=&\gamma_s\equiv 1-\rho_{o}(\alpha)\qquad {\rm (shockening)}\\
  \gamma&=&\gamma_d\equiv \rho_{o}(\alpha)\qquad {\rm (dual)}
\end{eqnarray}
The critical point $(\alpha_c,\gamma_c)$, being the point of
intersection of the two lines, has $\gamma_c=1/2$, and $\alpha_c$
satisfying Eq. (\ref{eq:15}) with $\rho_o=\gamma_c$.

The shape of the transition curve near $(\alpha_c,\gamma_c)$ for any
$r>0$ follows from Eq. \eqref{eq:15}, by noting that $dg(\rho)/d\rho$
is zero at $\rho=\gamma_c$ but not at $\rho=\alpha_c$.
This shows that the shape exponent $\chi$ is given by\cite{smsmb}
\begin{equation}
\label{eq:14}
\gamma-\gamma_c\sim \mid \alpha-\alpha_c\mid^{\chi},  \quad {\rm
with}\  \chi=1/2,
\end{equation}
close to the critical point.  The phase diagram is shown in Fig.
\ref{fig:del}a.  The critical point $(\alpha_c(r),\gamma_c)$ changes
with $r$ but the exponents remain the same as in the ASEP case.
{We note in passing that for $\gamma<\gamma_c$, one crosses
  the dual phase boundary so that there is a depletion layer near
  $x=1$. This layer is non-shockening in the sense that it never
  reaches saturation\cite{smvm,smb_jpa07}.  It acts as a shield for
  the boundary density, rendering the ``effective'' boundary condition
  at $x=1$ as $\gamma=\gamma_c$.  Consequently, thanks to this
  depletion layer, there is a continuation of the critical point in
  the phase diagram.}

The situation is completely different for $r<0$.  We refer to
Fig. \ref{fig:jrho}).  Let us start with
small injection rates.  There is a dual line at
$\gamma=\rho_o(\alpha)$ so long $\rho_o(\alpha)<\rho^{*}$ (Fig.
\ref{fig:jrho}).  There will be a symmetric shock centered around
$\rho=1/2$.  The transition to the shock phase (or the shockening
transition) is at $\gamma=\gamma_s\equiv 1-\rho_o(\alpha)$.  However,
with change of $\alpha$ when $\rho_o$ exceeds $\rho^*$, the boundary
layer bridges $\rho_o$ to $\rho_1<1/2$.  The
dual line continues to be $\gamma=\gamma_d=\rho_o(\alpha)$, but the
shock line is given by
\begin{equation}
  \label{eq:22}
  \gamma=\gamma_{s1}\equiv \rho_1(\alpha).
\end{equation}
The shock is therefore smaller in height, lying below $\rho=1/2$.  A
zero height shock at the boundary is now possible if the following two
conditions are met: {\it(i)} the injection rate $\alpha$ is such that
the outer solution gives the peak density at the boundary, i.e.,
$\rho_o(\alpha)=\rho_c$, and {\it (ii)} the withdrawal rate is such
that the accumulated density $\gamma$ is also $\rho_c$.  Under these
conditions, the density will then have an infinite slope at the $x=1$
boundary.  Therefore, $\gamma=\rho_c=\rho_o(\alpha)$ gives a critical
point.  This locates an off-center critical point C$_1$ in Fig
\ref{fig:del}c, with
\begin{equation}
  \label{eq:23}
\gamma_{c1}=\rho_c\  {\rm and} \ \alpha=\alpha_{c1},  
\end{equation}
where $\alpha_{c1}$ satisfies Eq. \ref{eq:15} with
$\rho_o=\gamma_{c1}$.

\figrhox

The density profile close to this critical point $C_1$ (for $r<0$) is
shown in Fig. \ref{fig:rhox}a. These are obtained by a numerical
solution of Eq.  \eqref{eq:4} in the limit of small $\epsilon$. For
$\frac{1}{2}>\gamma >\gamma_{c1}$, the shock height starts increasing
with $\gamma$ until the shock connects $\rho=\rho^*$ and $\rho=1/2$.
This is the largest shock one can produce from the left peak of the
current. For a range of $\gamma$ there is the possibility of a
downward shock.  One sees an overcrowded region sandwiched between the
$\alpha$-phase and the $\gamma$-phase, even though both $\alpha$ and
$\gamma$ are less than $1/2$, but the shock is centered around $1/2$.
The downward shock shifts towards $x=1$ as $\gamma$ is changed and it
disappears as it hits the boundary.  At this point the downward shock
gets converted into a depleted boundary layer.  With increase in
$\gamma$, the depleted layer undergoes a boundary transition to an
accumulated layer leading to a second shock from the second peak.
There is no obvious symmetry relation between the two upward shocks
except for their centers.  The sequence is shown in Fig.
\ref{fig:rhox}a.  For $\gamma>1 -\rho^*$, the two shocks merge into
one bigger one centered around $\rho=1/2$.

Next, consider $\alpha >\alpha_{c1}$ as shown in Fig. \ref{fig:rhox}b.
With the largest shock from the low density peak of the current, it is
possible for the density to reach the second peak of the current,
$\rho=1-\rho_c$, at $x=1$.  This gives the second critical point
because the shock height starts from zero as $\gamma$ exceeds
$1-\rho_c$.  The second critical point C$_2$ at
$(\alpha_{c2},\gamma_{c2})$ is determined by
\begin{subequations}
\begin{eqnarray}
  \label{eq:12}
  \gamma_{c2}&=&1-\rho_c, \\
   g(\gamma_{c2})&=&2\Omega\ (1-x_s) +g(0),\\
   g(0)         &=& 2\Omega\ x_s +g(\alpha_{c2}),
\end{eqnarray}
\end{subequations}
where $x_s$ is the position of the first shock, and $g$ is from Eq.
(\ref{eq:7}) with $r<0$.  Fig.  \ref{fig:del}c shows the nature of the
phase boundary for a given value of $r<0$.  These two critical points
$C_1, C_2$ merge as $r\to 0$.  Given the form of $g(\rho)$ in Eq.
\eqref{eq:6}, we find
\begin{equation}
  \label{eq:13}
  \mid\gamma_{c1}-\gamma_J\!\mid\;  \sim \;
  \mid\gamma_{c2}-\gamma_J\!\mid\;  \sim \;{\mid \!r\!\mid}^{1/2}, \ (r\to 0-),
\end{equation}
where $\gamma_{\rm J}$ is the critical value at $r=0$.  The power law
dependence is expected to be universal.  The bifurcation at point J is
shown in Fig. \ref{fig:pntJ}a in the projected plane of $\gamma$-$r$.
The full three dimensional lines\footnote{{The topology of
    the phase diagram around J is different from the topology for the
    equilibrium tricritical point\cite{ahar}.  For the latter, the
    curves meet tangentially but here the post-bifurcation curves have
    infinite slope at J.}}  are shown in Fig. \ref{fig:pntJ}b.  An
analysis as done for the $r>0$ case shows that the shape of the
boundaries at these new critical points would be similar to that for
the $r>0$ case, i.e the shape exponent will be $\chi=1/2$ as in Eq.
\eqref{eq:14}, implying universality.

Let us now consider the special point J.  The physical picture
developed above remains valid here. The tricritical point J, being at
the intersection of the shock line and the dual line, is at
\begin{equation}
\label{eq:10}
g_{r=0}(\gamma_{\rm J})=2\Omega + g_{r=0}(\alpha_{\rm J}),\quad
\gamma_{\rm J}=1/2. 
\end{equation}
Since $S_1(\rho)\sim (\rho- 1/2)^3$, the shape exponent is 
\begin{equation}
\label{eq:9}
\chi_{\rm J}=1/4,
\end{equation}
where $\chi_{\rm J}$ is defined as 
\begin{equation}
  \label{eq:19}
  \gamma-\gamma_{\rm J}\sim \mid \alpha-\alpha_{\rm J}\mid^{\chi_{\rm J}}.
\end{equation}
The phase boundary is shown in Fig. \ref{fig:del}b.  The first order
transition lines as shown in Fig. \ref{fig:del} become surfaces in the
three dimensional phase diagram. These first-order surfaces end on
critical lines and these critical lines are shown in Fig.
\ref{fig:pntJ}b.  The locus of the intersection of the first order
lines for $r<0$ (triangle in Fig \ref{fig:del}c) in the extended phase
diagram is a line ending at J.  This is the line on which both the
shocks connecting $\rho=\rho^*$ to $\rho=1/2$ and $\rho=1/2$ to
$\rho=1-\rho^*$ are just at $x=1$. This completes the identification
of J as the tricritical point where a first order line is converted
into a continuous one.

The crossover from the tricritical to the critical behaviour is
obtained by expanding Eq. \ref{eq:15} around $\alpha=\alpha_J,
\rho_o=\gamma_J$ and $r\to 0+$.  The shape of the transition curves
and also the height of the shocks near the tricritical point can be
described by a scaling form
\begin{equation}
\label{eq:11}
\alpha -\alpha_{\rm J} \sim\; \mid\! \gamma-\gamma_{\rm
J}\!\mid^{1/\chi_{\rm J}}\  {\cal F}\left( 
  {r}\;{ \mid\! \gamma-\gamma_{\rm J}\!\mid^{-\psi_{\rm
J}}} \right),
\end{equation}
where ${\cal F}(x)$ is a scaling function with the crossover exponent
$\psi_{\rm J}=2$.  For $x\to 0$, ${\cal F}(x) \to$ constant while for
large $x$, ${\cal F}(x)\sim x$.  For $r\neq 0$, with $\gamma$ close to
the critical value, the scaling variable becomes large and one
recovers the critical value $\chi^{-1}=\chi^{-1}_{\rm J}-\psi_{\rm
  J}=2$.  This also shows the importance of the tricritical point
because it controls the scaling behaviour in its neighbourhood.
 
Within the window of the single shock phase for $r<0$, there is a
region of downward shock bounded by critical and first order lines.
The peculiarity of the downward shock is that it does not traverse the
whole lattice but goes critical within the track.  The thickness of
the region in the phase diagram is determined by the shallowness of
the minimum at $\rho=1/2$.  Details of these lines and the region will
be reported elsewhere.

In summary, we have shown that the repulsive interaction induced
particle-hole symmetry breaking in the form of a double peaked current
in one dimensional asymmetric exclusion process leads to a
nonequilibrium tricritical point.  This tricritical point controls the
scaling behaviour in its neighbourhood.  The tricritical exponents are
different from the critical ones and an additional crossover exponent
is required for the crossover to the critical behaviour.  The phase
diagram also shows a region of downward shock.  It remains to be seen
if this is a necessity for the nonequilibrium tricriticality.

 \acknowledgments

The authors acknowledge the support of Saha Institute of Nuclear
Physics (SINP), Kolkata, where part of the work was done.  JM was supported
by CAMCS, SINP, during her work at SINP.

\end{document}